\documentclass[twocolumn]{aastex6}

\def \Msol{{\rm M}_{\odot}}
\def \logm{\log(m/\Msol)}
\newcommand{\myrefsec}[1]{\hyperref[#1]{Section~\ref*{#1}}}

\begin{document}


\title{Constraints on Quenching of  $z\lesssim2$ Massive Galaxies from the Evolution of the Average Sizes of  Star-Forming and Quenched Populations in COSMOS}


\author{
A. L. Faisst\altaffilmark{1,2},
C. M. Carollo\altaffilmark{1},
P. L. Capak\altaffilmark{2,3},
S. Tacchella\altaffilmark{1},
A. Renzini\altaffilmark{4},
O. Ilbert\altaffilmark{5},
H. J. McCracken\altaffilmark{6},
N. Z. Scoville\altaffilmark{3}
}


\affil{$^{1}$Institute for Astronomy, Swiss Federal Institute of Technology, 8093 Z\"urich, Switzerland}
\affil{$^{2}$Infrared Processing and Analysis Center, California Institute of Technology, Pasadena, CA 91125, USA}
\affil{$^{3}$Cahill Center for Astronomy and Astrophysics, California Institute of Technology, Pasadena, CA 91125, USA}
\affil{$^{4}$NAF - Osservatorio Astronomico di Padova, Vicolo dell'Osservatorio 5, I-35122 Padova, Italy}
\affil{$^{5}$Aix Marseille Universit\'e, CNRS, LAM (Laboratoire d'Astrophysique de Marseille) UMR 7326, 13388, Marseille, France}
\affil{$^{6}$Institut d'Astrophysique de Paris, CNRS \& UPMC, UMR 7095, 98 bis Boulevard Arago, 75014, Paris, France}


\email{afaisst@ipac.caltech.edu; Twitter: @astrofaisst}




\begin{abstract}
We use $>$9400 $\logm$$>$10  quiescent and star-forming galaxies at $z\lesssim2$ in COSMOS/UltraVISTA to study the average size evolution of  these systems, with focus on the rare, ultra-massive population at $\logm>11.4$. The large 2-square degree survey area delivers  a sample of $\sim400$ such ultra-massive systems. Accurate sizes are derived using a calibration based on high-resolution images from the Hubble Space Telescope. We find that, at these very high masses, the size evolution of star-forming and quiescent galaxies is almost indistinguishable in terms of normalization and power-law slope.
We use this result to investigate possible pathways of quenching massive $m>M^*$ galaxies at $z<2$. We consistently model the size evolution of quiescent galaxies from the star-forming population by assuming different simple models for the suppression of star-formation. These models include an instantaneous and delayed quenching without altering the structure of galaxies and a central starburst followed by compaction. We find that instantaneous quenching reproduces well the observed mass-size relation of massive galaxies at $z>1$. Our starburst$+$compaction model followed by individual growth of the galaxies by minor mergers is preferred over other models without structural change for $\logm>11.0$ galaxies at $z>0.5$.
None of our models is able to meet the observations at $m>M^*$ and $z<1$ with out significant contribution of post-quenching growth of individual galaxies via mergers.
	We conclude that quenching is a fast process in galaxies with $ m \ge 10^{11} M_\odot$, and that major mergers likely play a major role in the final steps of their evolution. 
\end{abstract}



\keywords{galaxies: evolution -- galaxies: structure -- galaxies: fundamental parameters}


\section{Introduction}


    
Quiescent (or quenched) galaxies -- here defined to be galaxies that have heavily-suppressed specific star-formation rates (specific SFRs) relative to the star-forming ``main sequence'' \citep[e.g.,][]{DADDI07,NOESKE07} -- host about half of the mass in stars in the local Universe \citep{BALDRY04}, and have been observed in substantial numbers as early as $z\sim2$ \citep[e.g.,][]{ILBERT13,MUZZIN13,DAVIDZON17}.
Understanding the dominant processes responsible for the shut-down of their star-formation (often referred to as "quenching"), as well as the connection between these processes and galaxy structure are key for understanding the evolution of the whole galaxy population over cosmic time.

		Suppressed specific SFRs are not the only intriguing property of quiescent galaxies. At least in terms of 'light', these have on average substantially larger spheroidal components and smaller half-light radii ($R_{e}$) compared to their star-forming counterparts at a given stellar mass and redshift \citep[e.g.,][]{SHEN03,SZOMORU12,CIBINEL13}. At least in part this size difference is likely contributed by 'nurture', in particular post-quenching 'fading' of stellar populations at large radii \citep[][]{TACCHELLA15b,CAROLLO16}, but it is also possible that part of the difference may be imprinted by 'nature', i.e., different formation processes for spheroids and disks. Also intriguing is that the population-averaged sizes of quiescent galaxies of a given mass has increased by a factor of $\sim3$ since $z=2$. This average size growth is similar to that of star-forming disk galaxies, which are expected and observed to increase their individual (disk) sizes more or less proportional to $(1+z)^{-1}$, through continuous accretion of gas from their halos \citep[e.g.,][]{MO98,OESCH10,MOSLEH12,NEWMAN12,HUANG13,SHIBUYA15,MOSLEH17}. Individual quiescent galaxies form however by definition no new stars, and thus their only channel for individual mass and size growth is provided by gas-poor mergers. Averaging over a large mass range, several studies suggest indeed that mergers are  important contributors for the size growth of quiescent galaxies \citep[e.g.,][]{TOFT07,BUITRAGO08,FRANX08,STOCKTON08,KRIEK09,WILLIAMS10,NEWMAN12,OSER12,WHITAKER12,BELLI14,BELLI15}. 
		
		Analyses  in thinner bins of stellar mass suggest however a threshold mass -- roughly around  $m~<~{\rm M^{*}}~\sim~10^{11}~M_\odot$, the characteristic mass of the Schechter \citep[][]{SCHECHTER76} fit to galaxy mass functions\footnote{The value of $M^*$ is remarkably constant for star-forming and quiescent galaxies and at all epochs since $z\sim4$; \citep[see e.g.][]{ILBERT13,MUZZIN14,DAVIDZON17}} -- below and above which different mechanisms may be responsible for the average size growth of quiescent galaxies. In particular, at $m \lesssim M^{*}$, a number of studies indicate that the growth in average size of the quiescent population is dominated  by the addition  of larger galaxies at later times, as a result of the continuos addition of newly quenched galaxies to the large-size-end of the size function \citep{CAROLLO13,CASSATA13,SARACCO14}. This picture is substantiated by the stellar ages of compact (older) and large (younger) quiescent galaxies at a given stellar mass and epoch \citep[][]{SARACCO11,ONODERA12,CAROLLO13,BELLI15,FAGIOLI16,WILLIAMS16}. 
		It is only above $M^{*}$ that dissipationless mergers are expected to be important \citep[e.g.,][]{PENG10} and all studies of galaxy sizes indeed agree on them playing the dominant  role in leading to the growth of individual quiescent galaxies in mass and size  \citep{CAROLLO13,POGGIANTI13,BELLI14,BELLI15}. 
		
	The above results may indicate that different quenching mechanisms could be at work below and above $M^{*}$. Theoretically, there are many candidate mechanisms for quenching \citep[e.g.,][]{BIRNBOIM03,CROTON06,BIRNBOIM07,BOURNAUD07,KAWATA08,MARTIG09,PENG10,FELDMANN11,DELUCIA12,HEARIN13,CEN14a,DEKEL14,MANDELKER14,SCHAYE15,TACCHELLA16a,TACCHELLA16b}, and identifying observationally the correct ones is a non-negligible challenge -- not least since, as discussed in \citet{CAROLLO13a} and demonstrated in \citep[][]{LILLY16}, correlations of observed quantities do not necessarily indicate a causal relation between them. 

	Different quenching mechanisms are expected to act on different time scales and result in different morphological transformations of galaxies. Therefore, constraining these is an important step towards understanding the dominant processes that lead to galaxy quiescence in these populations.
	For example, the cut-off of gas inflow onto a star-forming galaxies is expected to lead to the exhaustion of star-formation over long timescales, which are set by the time needed for star formation to consume the gas reservoir of a galaxy. It is likely that the star formation ceases smoothly over the galaxy's disk thereby not significantly changing its observed morphology. In contrast, a gas-rich major merger might lead to a starburst and thus to a fast consumption of gas on dynamical timescales of order $100-200~{\rm Myrs}$. Furthermore, a substantial change in the morphology of the galaxies is expected with an apparent compaction in light induced by the centrally confined starburst \citep[e.g.,][]{BARRO13,ZOLOTOV15}.
	A number of studies have focused their attention on quenching timescales and their dependence on galaxy properties. For relatively massive galaxies, and in particular satellites in groups and clusters at low redshifts, there is growing evidence that the transition from active star-formation to quiescence takes of order $2-4~{\rm Gyrs}$ \citep{VONDERLINDEN10,DELUCIA12,CIBINEL13a,MOK13,TRINH13,WETZEL13,HIRSCHMANN14,MUZZIN14,SCHAWINSKI14,TARANU14,PENG15}.  At redshifts of order $z\sim2$ and for massive galaxies,  \citet{TACCHELLA15b,TACCHELLA16b} show that suppression of star-formation starts at the center of galaxies and slowly progresses outwards on timescales of $1-3~{\rm Gyrs}$. It is however unclear at this point whether the observed centrally-suppressed specific SFRs are the outcome of a ``quenching mechanism'' \citep[e.g., through central gas and stellar ``compaction''; ][]{DEKEL14,TACCHELLA16a} or the natural outcome of inside-out galaxy formation \citep[e.g,][]{LILLY16}.

In this paper we make a new attempt to constrain the processes that quench massive, $m>M_*$ star-forming galaxies at $z<2$ via studying the timescales and morphological changes using as diagnostic tool the size evolution of both the star-forming and quiescent galaxy populations at $z<2$. We will show that this further enables us to set constraints on the amount and properties of mergers in this massive population

It is now well established that studies of $z\sim2$ galaxies crucially  need imaging in the near-infrared in order to measure their rest-frame optical properties (especially sizes); the near-infrared images need to be furthermore  quite deep in order to detect the faint and fading stellar populations of quiescent systems.  Much progress has been made using data from the HST CANDELS survey \citep{GROGIN11,KOEKEMOER11}. Very massive galaxies are however rare, and increasingly so at increasingly higher redshifts \citep[with number densities of less than $10^{-4}$ per ${\rm Mpc}^{3}$ at $z>1$;][]{ILBERT13}. Assembling a sufficiently large number of such galaxies to enable a statistical study requires imaging over a  large area of sky. With its two square-degree area coverage, the Cosmological Evolution Survey \citep[COSMOS\footnote{\url{http://cosmos.astro.caltech.edu/}},][]{SCOVILLE07} enables us to assemble a sample of more than 400 ultra-massive galaxies (UMGs, $\logm > 11.4$) in the redshift range $0.2 < z < 2.5$. Another advantage of COSMOS is  its $>30$ pass-band coverage from  UV to IR wavelengths, which enables the derivation of very accurate stellar masses and photometric redshifts. Last but not least, the deep near-IR data of the UltraVISTA survey on COSMOS \citep{MCCRACKEN12,LAIGLE16} allow an accurate separation of star-forming and quiescent galaxies across this entire redshift range \citep[e.g.,][]{ILBERT13}.
	A drawback of the UltraVISTA data is their seeing-limited Point Spread Function (PSF), which full width at half maximum (FWHM) is typically about $0.8"$, and therefore hampers the measurement of reliable galaxy sizes. To overcome this limitation, we correct the UltraVISTA size measurements using as a calibration reference the $\sim3\%$ of the COSMOS area that is covered by the CANDELS/COSMOS legacy survey.

The paper is organized as follows.  In \myrefsec{sec:data} we describe the  datasets that we have used in this work. In \myrefsec{sec:sample} we describe the selection criteria for separating star-forming and quiescent UMGs, and in \myrefsec{sec:size}  the procedure that we have followed to measure the galaxy sizes. In the same Section we also present the calibration of the UltraVISTA sizes that we have performed using the HST CANDELS size measurements for the $\sim9000$ galaxies for which both datasets are available. The final (calibrated) size measurements are presented and discussed in \myrefsec{sec:results}. In \myrefsec{sec:model} we present our model that we use to predict the average size evolution of quiescent galaxies through the redshift range of our analysis.
The model predictions are compared with the observed size evolutions in \myrefsec{sec:discussion}, where we furthermore describe the additional modifications to the predicted trends that are introduced by galaxy mergers. We summarize our main results in \myrefsec{sec:ending}.

Note that: all magnitude are given in the AB system \citep{GUNN86}; stellar masses ($m$) are scaled to a \cite{CHABRIER03} initial mass function (IMF); we assume a flat cosmology with $\Omega_{\Lambda}=0.7$, $\Omega_{m}=0.3$, and $H_{0}~=~70$~km~s$^{-1}$~Mpc$^{-1}$.

\section{Data}\label{sec:data}


\subsection{UltraVISTA near-IR imaging data}
	
As mentioned in the previous section, near-IR data on a large area is crucial for the study of massive galaxies at high redshifts.
	Therefore, the backbone of this work is the UltraVISTA survey carried out on the 4.1 meter Visible and Infrared Survey Telescope for Astronomy (VISTA) located at the Paranal observatory in Chile. This survey covers $1.5~{\rm deg}^{2}$ of the COSMOS field in the near-infrared bands $Y$, $J$, $H$, and $K_{s}$.
	Specifically, we use the (unpublished) UltraVISTA data-release (DR) 2 imaging data. Compared to DR1, this release has an improvement in $H$-band by up to 1 magnitude in the ultra-deep stripes (covering roughly 50\% of the field) and $\sim0.2$ magnitudes on the deep stripes. The typical exposure times per pixel are between 53 and 82 hours, leading to $5\sigma$ sensitivities of 25.4AB, 25.1AB, 24.7AB, and 24.8AB in $Y$, $J$, $H$, and $Ks$ band within $2\arcsec$ aperture.
    The reduction of the imaging data is similar to DR1 \citep[see][]{MCCRACKEN12} and is briefly outlined in the following: the data was taken in three complete observing seasons between December 2009 and May 2012. The individual science frames are visually inspected to remove bad frames (e.g., due to loss of auto-guiding). Each frame is sky subtracted before stacking which leads to a very flat combined image with a very small variation in background flux. The combined frames have an \textit{average} $H$-band seeing of $0.75\arcsec \pm 0.10\arcsec$. The final photometric calibration is done by using non-saturated stars from the Two Micron All Sky Survey \citep[2MASS;][]{SKRUTSKIE06} sample leading to an absolute photometric error of less than 0.2 magnitudes.

\subsection{Photometric redshift and stellar mass catalog}

Our galaxy selection (see below) is based on the public COSMOS/UltraVISTA catalog in which galaxies are selected from a combined $YJHK_{s}$ image \citep{ILBERT13}. This has advantages compared to purely optical selected catalogs as it more sensitive to galaxies with red colors, e.g., dusty star-forming galaxies or quiescent galaxies with old stellar populations. The catalog comprises photometric redshifts, stellar masses, and other physical quantities derived from SED fitting on $>30$ pass-bands from UV to IR (PSF homogenized) for more than $250,000$ galaxies on COSMOS \citep[see e.g.,][]{CAPAK07,ILBERT13}.
	The photometric redshifts in that catalog are derived using \texttt{Le Phare} \citep{ARNOUTS02,ILBERT06} employing different templates including a range of galaxy types from elliptical to young and star-forming. These redshifts have been verified to have a precision of $\sigma_{\Delta z/(1+z)} = 0.01$ up to $z=3$ by comparison to a sample of more than $\sim10.000$ spectroscopically confirmed star-forming and quiescent galaxies. Physical quantities (mass, SFR, etc) are fitted by \texttt{Le Phare} at fixed photometric redshift using a library of synthetic composite stellar population models based on \citet{BRUZUALCHARLOT03}. These models include different dust extinctions (following a \citet{CALZETTI00} dust extinction law), metallicities, and star formation histories (following exponentially declining $\tau$ models). Also, emission line templates are included. The emission line flux is derived from the observed UV light using empirical relations. All these parameters have been verified by a number of other fitting routines including \texttt{ZEBRA} \citep{FELDMANN08} and its upgraded version \texttt{ZEBRA+} \citep{OESCH10a,CAROLLO13}.
	The typical uncertainties in masses are on the order of $0.3~{\rm dex}$. All quantities are computed for a \citet{CHABRIER03} IMF. The stellar masses are defined as the integral of the star-formation histories of the galaxies, thus representing the total galaxy mass of a galaxy rather than its mass in active stars. In the following, the stellar masses quoted by other studies are converted to total masses if necessary. These corrections, calculated using \citet{BRUZUALCHARLOT03} models with solar metallicity and exponentially declining as well as constant star formation histories, can be up to $0.2\,{\rm dex}$ for quiescent galaxies with ages of 1 billion years and above, while they are less substantial for star-forming galaxies.

\subsection{CANDELS/COSMOS near-IR imaging data}
To calibrate the sizes measured on the ground based UltraVISTA imaging data, we make use of the overlap between UltraVISTA and HST based CANDELS/COSMOS survey \citep{GROGIN11,KOEKEMOER11}. 
The latter covers $0.06{\rm deg}^{2}$ on sky (roughly 1/25th of the total UltraVISTA field) in the WFC3/IR F160W pass-band, similar to the UltraVISTA $H$-band, however at a much high resolution (more than 8 times smaller PSF).
We use of the latest publicly available data release of the COSMOS/F160W mosaic (by February 2013) with a total exposure time of 3200s and a sensitivity of 26.9 AB (5$\sigma$ for a point source).


\begin{figure}
\centering
\includegraphics[width=1.0\columnwidth, angle=270]{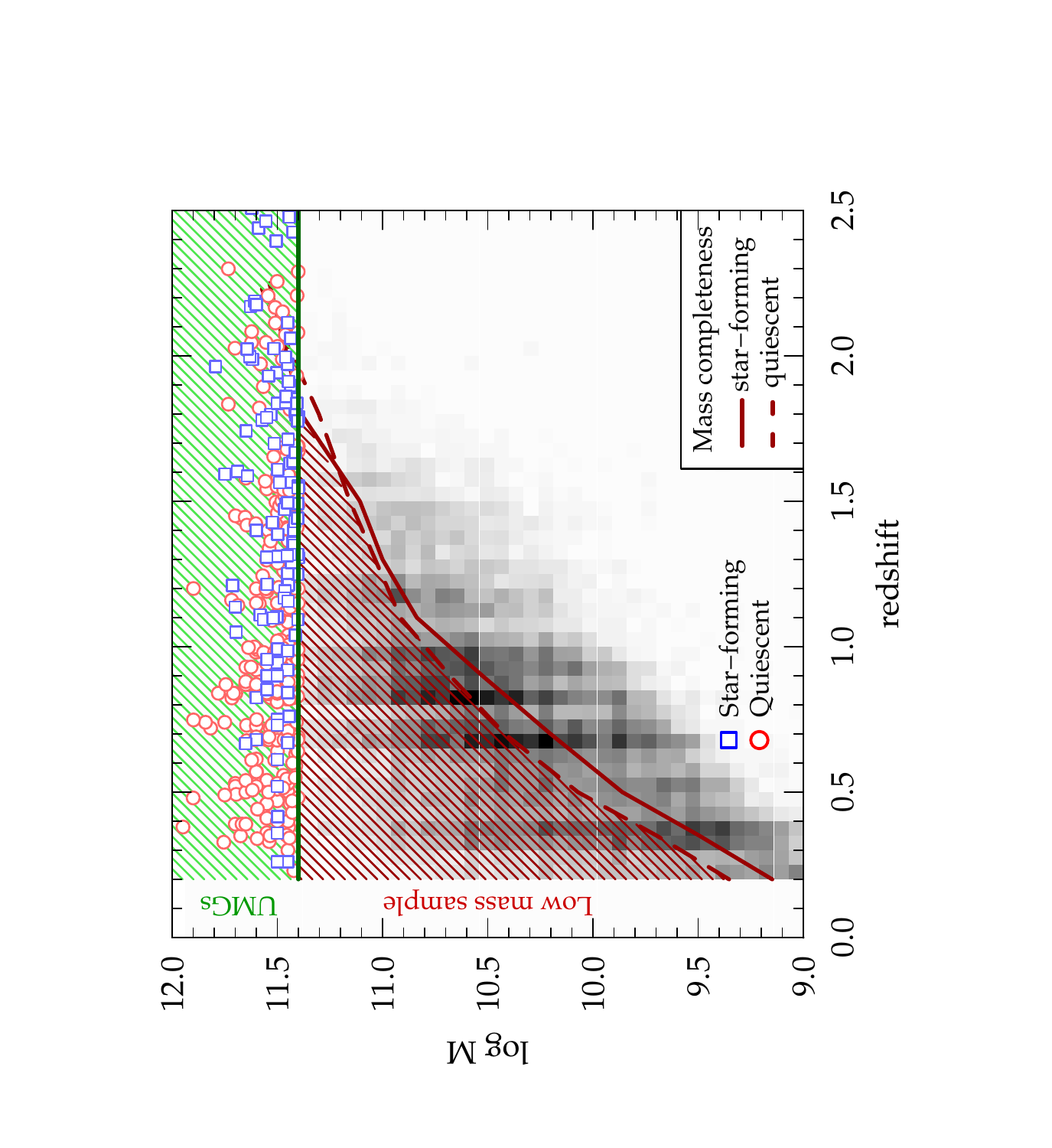}
\caption{Sample selection. Our sample of UMGs ($\log(m/M_{\odot}) >11.4$) at $0.2 < z < 2.5$ is shown with large symbols. The sample is split into quiescent (red circles) and star-forming (blue squares) according to their location on the rest-frame $(NUV - r)$ vs. $(r - J)$ diagram (see also \autoref{fig:sfqu}). Other galaxies with $\log(m/M_{\odot}) < 11.4$ and $H<21.5$AB are shown in gray. The dark red line shows the 90\% mass completeness for star-forming (solid) and quiescent (dashed) galaxies as described in the text. The low mass sample therefore consists of the galaxies shown in gray scale which are in the dark red hatched region.
\label{fig:sample}}
\end{figure}

\begin{figure*}
\begin{center}
\includegraphics[width=0.8\textwidth, angle=270]{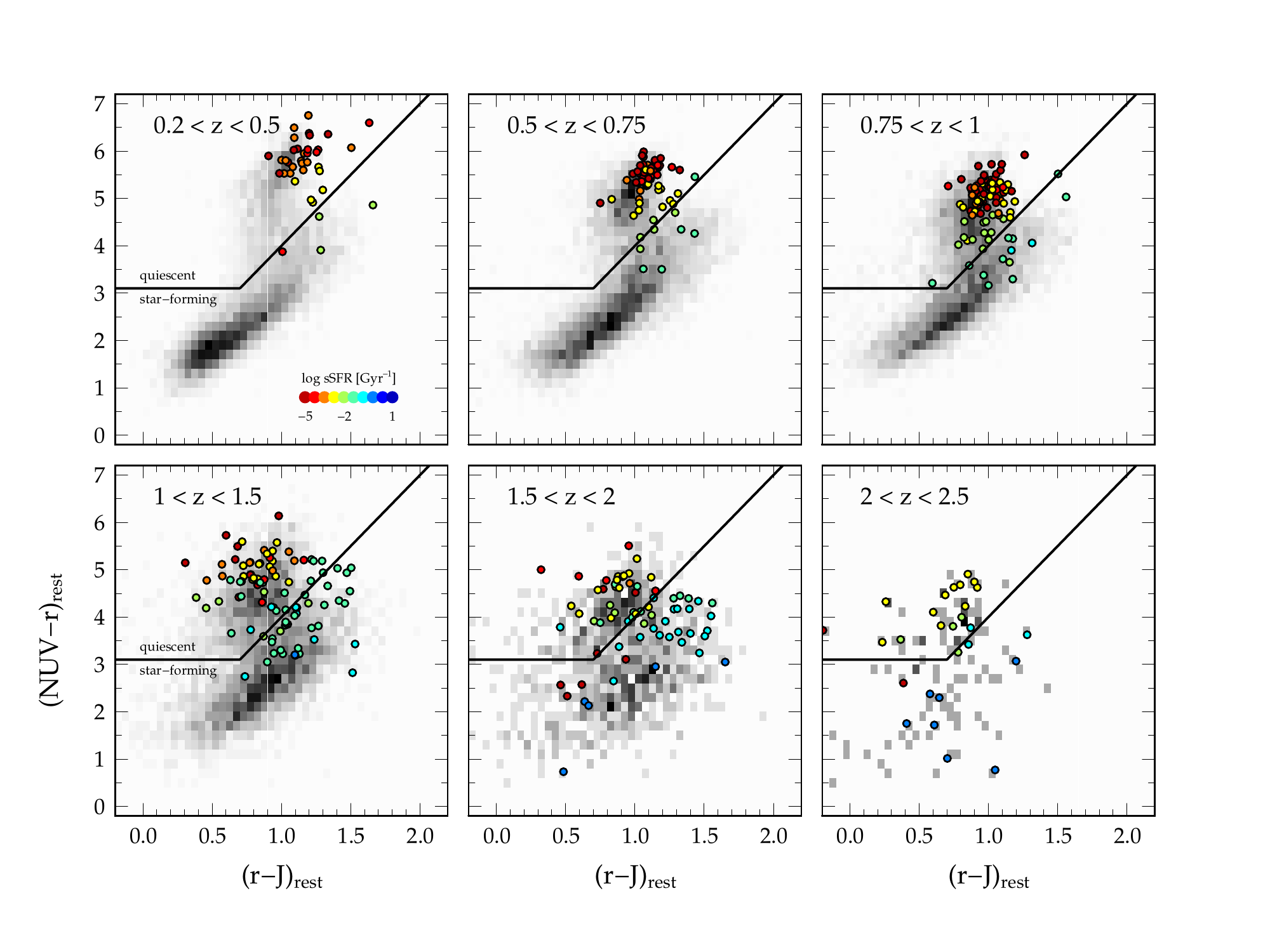}
\caption{Selection of quiescent (upper left of solid line) and star-forming (lower right of solid line) UMGs on the rest-frame $(NUV - r)$ vs. $(r - J)$ diagram for 6 different redshift bins. The UMGs are color coded by their sSFR. This shows that the color-color cut efficiently separates quiescent galaxies with $\log({\rm sSFR}/{\rm Gyr}) \sim -1$ to $-2$. The gray background shows less massive galaxies with $H<21.5$AB in the same redshift bins.
\label{fig:sfqu}}
\end{center}
\end{figure*}

\section{The Sample}\label{sec:sample}

In the following, we describe the selection of massive galaxies at $\logm > 11.4$ building our main galaxy sample as well as less massive galaxies ($10.0 < \logm < 11.4$) that we use for the calibration of the ground-based size measurements. Furthermore, we split this sample into quiescent and star-forming galaxies.

\subsection{High- and low-mass galaxies}\label{sec:galselection}
The selection of the high- and low-mass galaxy sample is based on the near-IR COSMOS/UltraVISTA photometric catalog (as described above), which allows for the selection of dusty star-forming and quiescent galaxies.

	We select a total sample of 403 massive galaxies satisfying $\log(m/M_{\odot}) > 11.4$ and $0.2 < z_{phot} < 2.5$ (green hatched region in \autoref{fig:sample}). We have verified these galaxies visually to be real (i.e., not artifacts or stars). The exact value of this mass limit has been chosen to correspond to the 90\% completeness limit at a $H$-band magnitude of $21.5~{\rm AB}$ at $z<2.5$, which allows us provide reliable size measurements for these galaxies (see \myrefsec{sec:size}). For the estimation of the mass completeness we have used the identical method as described in \citet{POZZETTI10}. With this mass cut, we select the most massive observable galaxies with a number density less than $10^{-4}$ Mpc$^{-3}$ and $10^{-5}$ Mpc$^{-3}$ at $z\sim0.5$ and $z\sim2$. These galaxies may be the progenitors of today's most massive galaxies, assuming these most massive galaxies keep their ranking through cosmic time. This is verified by more complicated methods of progenitor selections, including the selection of galaxies at a constant galaxy number density \citep{MARCHESINI14}, or using semi-empirical models that take into account galaxy mergers \citep{BEHROOZI13a}.

	The (mass complete) low-mass galaxy control/calibration sample is selected in a similar way to have $10.0 < \log(m/M_{\odot}) < 11.4$ and $H<21.5~{\rm AB}$. The mass completeness limit at $H=21.5$AB as a function of redshift is shown in \autoref{fig:sample} by the red line (solid for star-forming and dashed for quiescent galaxies). The low mass control sample (9000 galaxies in total) is consequently selected to be above the combined completeness limit of the star-forming \textit{and} quiescent galaxies and satisfies three stellar mass bins of $10.0 < \logm < 10.5$, $10.5 < \logm < 11.0$, and $11.0 < \logm < 11.4$ with the corresponding redshift ranges $0.2 < z < 0.45$, $0.2 < z < 0.75$, and $0.2 < z < 1.25$.

\subsection{Selection of quiescent and star-forming galaxies}
    
    We split our sample into quiescent and star-forming galaxies by making use of the rest-frame $(NUV - r)$ versus $(r - J)$ color diagnostics \citep[see][]{WILLIAMS09,ILBERT10,CAROLLO13,ILBERT13}.
    In \autoref{fig:sfqu} we show the rest-frame $(NUV - r)$ versus $(r - J)$ diagram for six different redshift bins with our main sample of massive galaxies. The black line in each panel divides the quiescent (upper left) from the star-forming (lower right) galaxy population. Our $\logm>11.4$ galaxies are shown with large symbols color coded by their specific star-formation rate (${\rm sSFR}~\equiv~{\rm SFR}/m$, the inverse of the mass doubling time scale) derived from SED fitting.
    All the other galaxies at lower stellar masses in the same redshift bin and $H<21.5$AB are shown in gray scale. We find that the color-color diagram efficiently isolates quiescent galaxies with $\log({\rm sSFR}/{\rm Gyr}) \sim -1$ to $-2$ (depending on redshift, as expected).
        We note that this color selection is very similar to the widely used $(U-V)$ versus $(V-J)$ selection but it is a slightly better indicator of the current versus past star formation activity \citep[e.g.,][]{MARTIN07,ARNOUTS07}. We have verified that other selections of quiescent and star-forming galaxies (e.g., by sSFR or $(U-V)$ versus $(V-J)$) do not change the results of this paper.

\section{Size measurements and calibration}\label{sec:size}

As we have already discussed in the introduction to this paper, we are investigating the quenching process in massive galaxies via the average size evolution of star-forming and quiescent galaxies. Reliable size measurements are therefore crucial. We denote with ''size'' the observed semi-major axis half-light radius, $R_{e}$.
	While we benefit from the large area of the COSMOS/UltraVISTA survey to select very massive galaxies, its poor resolution and PSF hampers the accurate measurement of galaxy structure parameters.
	
	In this section, we lead in detail through
	\textit{(i)} the determination of a spatially varying PSF,
	\textit{(ii)} the basic measurement of galaxy sizes, and
	\textit{(iii)} our 2-step size-calibration procedure using simulated galaxies and the HST based CANDELS imaging.
	Finally, we outline how we correct for the band-shifting across redshift in our sample.

\subsection{Determination of the spatially varying PSF}
Galaxy sizes are measured by the use of \texttt{GALFIT}, which takes into account the effect of PSF \citep{CHIEN10}.
Therefore, the understanding of the PSF size (full width at half maximum, FWHM), shape, and spatial variation is crucial.
    We represent the 2-dimensional PSF at a given position $(x,y)$ by a Moffat profile \citep{MOFFAT69}:

\begin{equation}
F(x,y)
= \frac{\beta - 1}{\pi\alpha^{2}}\left[1+\left(\frac{(x-\mu_{x})^{2}+(y-\mu_{y})^{2}}{\alpha^{2}}\right)\right]^{-\beta},
\end{equation}
where $\mu_{x}$, $\mu_{y}$, $\alpha$, and $\beta$ are free fitting parameters.
The FWHM of a PSF in this parametrization is given by

\begin{equation}
\textup{FWHM}(\alpha,\beta) = 2\alpha\sqrt{2^{1/\beta}-1}.
\end{equation}

This has been shown to be a good approximation for ground based PSFs and has the advantage over a pure Gaussian as it represents better the wings of the PSF \citep[e.g.,][]{TRUJILLO01}.
	In order to create a spatially comprehensive PSF map, we select unsaturated stars between $16~{\rm AB}$ and $21~{\rm AB}$ from the HST based COSMOS/ACS $I_{F814W}$-band catalog \citep{LEAUTHAUD07}. We select them according to their \texttt{SExtractor} stellarity parameter (larger than 0.9) and using diagnostic diagrams as color vs. color and magnitude vs. size. Furthermore, we inspect the stars visually and make sure that there are no close companion stars (or galaxies) visible on the ACS images.
	
	For each of these more than 3000 stars, we extract a $10\arcsec\times10\arcsec$ image stamp from the UltraVISTA $H$-band mosaic on which we will fit the PSF. We notice small shifts of the center of the stars between ACS and UltraVISTA data of a few tenths of arc seconds (likely caused by small differences in the coordinate systems, the large differences in the PSF size, and differences in the resolution of the images) which we correct for. We then fit the selected stars according to the above parametrization $F(x,y | \mu_{x}, \mu_{y}, \alpha, \beta)$. The accuracy and robustness of the fitting method was verified by generating stars with random FWHM between $0.2\arcsec < \textup{FWHM} < 1.2\arcsec$, add noise taken from real background images, and fit them in the same way as the real data. This test shows that we are able to recover the FWHM with an accuracy of better than $0.05\arcsec$. As a last cut, we require less than 5\% difference between the model and data in the enclosed flux up to 1.5 times the PSF FWHM. We end up with $\sim800$ PSF models across COSMOS/UltraVISTA. The PSFs show variations in their FWHM between $0.65\arcsec$ and $0.80\arcsec$. We assign to each galaxy an average PSF model created from the stars within $6\arcmin$, which we use for \texttt{GALFIT}.

\subsection{Guess-parameters for surface brightness fitting}\label{sec:sex}
In this section, we describe the determination of the initial values which are fed to \texttt{GALFIT}.
In order to have consistency between the initial values and the actual images on which we run \texttt{GALFIT}, we do not use the values given in the public COSMOS/UltraVISTA catalog, but we re-run Source Extractor (\texttt{SExtractor}, version 2.5.0, \citet{BERTIN96}) on the DR2 UltraVISTA $H$-band images.
	We run \texttt{SExtractor} with two different values of the \textit{DEBLEND\_MINCONT} for a better de-blending of galaxies next to brighter galaxies or stars. The \texttt{SExtractor} input parameters are tuned manually in order to optimize the source extraction. We mask \citep[using \texttt{Weightwatcher};][]{MARMO08} each star identified on the HST based COSMOS/ACS $I_{F814W}$-band images by a circle with a maximal radius $r_{\sigma}$ at which its flux decays to the background flux level. This maximal radius (which depends on the magnitude of the star) is determined by fitting $r_{\sigma}$ as a function of magnitude for a couple of different stars in a broad magnitude range. Furthermore, we match our catalog to the public UltraVISTA catalog and compare the measured magnitudes, which we find to be in excellent agreement.
	Finally, we extract each of our galaxies from our \texttt{SExtractor} catalog to use the measured galaxy position (\textit{X\_IMAGE} and \textit{Y\_IMAGE}), magnitude (\textit{MAG\_AUTO}), half-light radius (\textit{FLUX\_RADIUS}), axis ratio (ratio of \textit{A\_IMAGE} and \textit{B\_IMAGE}), and position angle (\textit{THETA\_IMAGE}) as initial parameters for \texttt{GALFIT}.

\begin{figure*}
\begin{center}
\includegraphics[width=1\textwidth, angle=0]{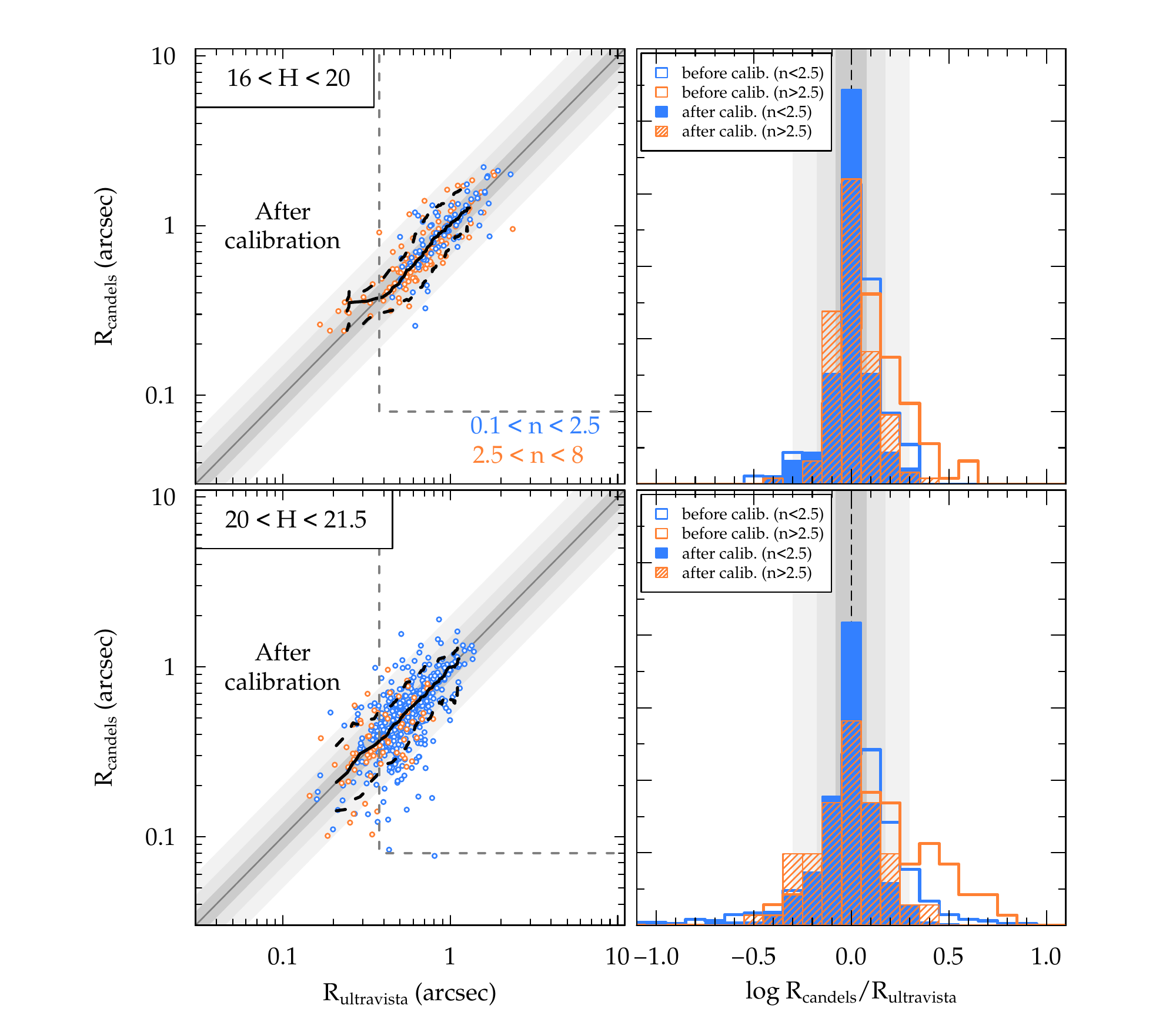}
\caption{Final calibration of sizes (in arc seconds) using the CANDELS imaging data. Shown are two different magnitude bins below $H=21.5$AB. The left panels show the comparison of semi-major half-light radii measured on CANDELS and ground-based UltraVISTA images after correction of systematic biases and calibration. Orange and blue points show galaxies with $0.1 < n <2.5$ and $2.5 < n < 8.0$, the gray regions show 20\%, 50\%, and 100\% discrepancies, and the black lines show a running median with scatter (dashed black lines). Furthermore, the dashed horizontal (vertical) line shows the CANDELS (UltraVISTA) PSF size.
The right panels show the normalized histograms of $\log (R_{candels}/R_{ultravista})$ for uncalibrated (empty) and calibrated (filled/hatched) UltraVISTA size measurements. Without calibrations, the UltraVISTA sizes are substantially under-estimated (tail towards large $\log (R_{candels}/R_{ultravista})$). This is true in particular for compact ($n>2.5$) galaxies. The histograms indicate an uncertainty for calibrated sizes of $\sim$50\%. The gray regions show 20\%, 50\%, and 100\% discrepancies.
\label{fig:calib}}
\end{center}
\end{figure*}

\subsection{Uncalibrated size measurements}\label{sec:uncalibmeas}
We use \texttt{GALFIT} to fit single Sersic profiles (parametrized by the half-light radius $R_e$, Sersic index $n$, total magnitude $M_{{\rm tot}}$, axis ratio $b/a$, and position angle $\theta$) to the observed surface brightness of our galaxies.
As described in the previous section, we use the \texttt{SExtractor} values measured on the DR2 COSMOS/UltraVISTA images as initial parameters. For the Sersic index, which is not known a-priori, we assume $n=2$ (and let it vary between $0 < n < 8$ during the fitting process).
The size of the image cutout on which \texttt{GALFIT} is run is variable between $71\times71$ and $301\times301$ pixels. The size is set to optimize the estimate of local sky background and to minimize the running time of \texttt{GALFIT} and is defined such that the cutout contains three times more sky pixels than pixels attributed to galaxy detections. Companion galaxies on the image cutout are fit simultaneously with the main galaxy if they are brighter than 25AB in $H$-band. All other detections of fainter objects are masked out and not taken into account in the $\chi^{2}$ minimization.
	To access the stability of the fits, we run \texttt{GALFIT} in two different configurations: In the first configuration (referred to as ``\textit{VARPOS}'') we let \texttt{GALFIT} fit the center of the galaxy within $\pm10$ pixels of the \texttt{SExtractor} input. In the second configuration (referred to as ``\textit{FIXPOS}'') we fix the galaxy position to its initial \texttt{SExtractor} value.
	
	We select good fits (either from the \textit{FIXPOS} or \textit{VARPOS} run) by comparing the results from the two configurations. We require that
\textit{(i)} $R_{e} > 0.1~{\rm px}$,
\textit{(ii)} the fitted position differ by less than $\sqrt{2}/2$ times the PSF FWHM from the \texttt{SExtractor} input
\textit{(iii)} the $R_{e}$ of the two configurations agree better than $50$\%, and 
\textit{(iv)} the total magnitude does not differ by more than 0.5 from the \texttt{SExtractor} total magnitude. 
	Roughly $70\%$ of our total sample galaxies satisfy these criteria and are used in the following for accessing the size evolution as a function of cosmic time.
	Due to their brightness and relatively large size, the above criteria result in a negligible cut for our massive $\logm>11.4$ but in principle could affect the following results and conclusions. We have investigated this in depth and find that mostly unresolved galaxies are affected by this without any clear relation with redshift. However, adding this small amount of galaxies to our sample at $\logm>11.4$ (keeping their small sizes as lower limits) impacts the median size versus redshift relations by less than $5\%$ compared to the general systematic uncertainties of the ground-based sizes of up to $50\%$. Furthermore, star-forming and quiescent galaxies are equally affected and therefore we do not expect significant impacts on our results.

\subsection{Correcting for measurement biases using simulated galaxies}
The measurement of galaxy structure is prone to many biases as discussed by several authors \citep{CAMERON07,CAROLLO13,CIBINEL13}. Small and compact galaxies are affected by the PSF (leading to an over-estimation of $R_{e}$); large and extended galaxies suffer surface brightness dimming in the outskirts (leading in under-estimation of $R_e$). Although \texttt{GALFIT} does take into account the effects of PSF and therefore partially cures these problems it has its limits.
It is therefore important to investigate possible biases and correct for them by using simulated galaxies. In the following, we outline this first step in our 2-step calibration process in more detail.

\subsubsection{Simulating galaxies}
We use \texttt{GALFIT} to create $\sim$1.5 million model galaxies on a grid in $(R_{e},M_{{\rm tot}},n,b/a)_{in}$ parameter space: $0.2 < n < 10$, $15~{\rm mag} < M_{{\rm tot}} < 26~{\rm mag}$, $0.2 < b/a < 1$, and $0.5 < R_{e} < 15$ pixels (corresponding to $0.075\arcsec < R_{e} < 2.250\arcsec$). The model galaxies are subsequently convolved with a PSF, equipped with Poisson noise, and added onto realistic sky backgrounds.
	For the latter, we account for the fact that the sky background noise ($\sigma_{{\rm sky}}$) varies across the COSMOS/UltraVISTA field by a factor $2$ or more (mainly between the deep and ultra-deep stripes). We compute $\sigma_{{\rm sky}}$ automatically in rectangles of $\sim 0.1\times0.1$ degrees across the field. For this end, we use the \texttt{SExtractor} catalog (see \myrefsec{sec:sex}) to mask out all detections and fit $\sigma_{sky}$ to the remaining non-masked pixels by assuming a Gaussian noise distribution. In order to make sure to remove all the light of galaxies and stars, we increase their semi-major and semi-minor axis as given by \texttt{SExtractor} by a factor of 10. We verify this procedure by manually measuring $\sigma_{{\rm sky}}$ at random positions.
	To take into account the variations in PSF and $\sigma_{\rm sky}$ we simulate galaxies in four different representations which will be interpolated in the end. We use two bracketing PSFs (FWHM = $0.65\arcsec$ and $0.85\arcsec$) as well as two bracketing $\sigma_{{\rm sky}}$ ($5.5\times10^{-6}$ and $2.0\times10^{-5}$ counts/s).
	On each of these model galaxies we run \texttt{SExtractor} and \texttt{GALFIT} in the same manner as for the real galaxies (as described in \myrefsec{sec:uncalibmeas}) to obtain $(R_{e},M_{{\rm tot}},n,b/a)_{out}$.
	This allows us to derive a correction function and discuss possible measurement biases as outlined below.

\subsubsection{Correction function}\label{sec:corrfunction}
We obtain a correction function, $\mathcal{S}(R_{e},M_{{\rm tot}},n,b/a)$, in an identical fashion as in \cite{CAROLLO13} and we refer the reader to this paper for additional details.
	We construct $\mathcal{S}$ such that it returns a 4 dimensional median correction vector $(\Delta R_{e},\Delta M_{{\rm tot}},\Delta n,\Delta b/a)$ for each point in measured $(R_{e},M_{{\rm tot}},n,b/a)_{{\rm meas}}$ parameter space. The median correction vector is constructed as the difference between the median of the 50 closest $(R_{e},M_{{\rm tot}},n,b/a)_{out}$ (with respect to $(R_{e},M_{{\rm tot}},n,b/a)_{meas}$) and the median of their true values $(R_{e},M_{{\rm tot}},n,b/a)_{in}$. We obtain this correction vector for each combination of PSF and $\sigma_{sky}$. The final correction vector is then obtained by an interpolation of the grid at the PSF and $\sigma_{sky}$ attributed to the galaxy for which the correction is computed.
	
	Because of our imposed magnitude cut of bright $H=21.5~{\rm AB}$, the correction in size (usually over-estimated) is on the order of less than $20\%$. The simulations also show that the detection rate of galaxies is $100\%$ in the worst case up to half-light sizes of at least $3\arcsec$ at $H=21.5~{\rm AB}$, corresponding to a surface brightness limit of $\sim25.2~{\rm mag}~{\rm arcsec}^{-2}$. This size corresponds to $\sim25~{\rm kpc}$ ($\sim20~{\rm kpc}$) at $z\sim2$ ($z\sim0.5$).
	
	The correction function allows an assessment of detection limits and a first correction for measurement biases. However, the simulated galaxies are ideal cases. The overlap between UltraVISTA and CANDELS is ideal to do a more thorough \textit{calibration} of our size measurement.

\subsection{Final calibration of size measurements using CANDELS}
The second step of our calibration process consists of the comparison of our measured (and corrected with $\mathcal{S}$) sizes with HST based structural measurements  on COSMOS/CANDELS, which has an overlap of $3\%$ with the central part of COSMOS. Because of the 2.5 times higher resolution and 4 times smaller PSF of the HST images, we consider the HST based size measurements to reflect the true galaxy sizes.
	We first measure the sizes of galaxies on the publicly available CANDELS F160W mosaic as these closest match the UltraVISTA $H$-band data. For this end, we use \texttt{SExtractor} in order to extract the sources and to get the initial parameters for \texttt{GALFIT} in the same manner as described above for the UltraVISTA based measurements.
	Subsequently, we run \texttt{GALFIT} for the extracted sources in the two configurations \textit{FIXPOS} and \textit{VARPOS} thereby applying the same selection criteria for good fits as described in \myrefsec{sec:uncalibmeas}. Furthermore, we apply the a correction function $\mathcal{S}$ as done before but with PSF and $\sigma_{{\rm sky}}$ matching those of the COSMOS/CANDELS images. In turn, we find corrections less than $5\%$ for galaxies at $H<21.5~{\rm AB}$. As a further check, we compare the size measurements to the public available COSMOS/CANDELS size catalog by \citet{VDW12} and find excellent agreement.

	The comparison between the HST-based ($R_{\rm candels}$) and ground-based ($R_{\rm ultravista}$) galaxy sizes and their calibration is shown in \autoref{fig:calib}. Shown are galaxies with Sersic indices $n<2.5$ (blue) and $n>2.5$ (orange) measured on the ground-based images in two magnitude bins at $H<21.5$AB (top and bottom row). Looking at the \textit{empty} histograms (showing the log-ratio of the sizes) on the right panels we see an under-estimation of $R_{{\rm ultravista}}$ by a factor 3 and more which we find to happen preferentially for galaxies smaller than the (UltraVISTA) PSF radius ($\sim 0.3\arcsec$) and with large Sersic $n$ (i.e., compact light distribution). Furthermore, an over-estimation of galaxy sizes preferentially happening for large galaxies ($R_{e} > 2\arcsec$) with small Sersic $n$.
	
	We calibrate our ground-based size measurements by constructing a calibration function $\mathcal{C}(R_{e},M_{{\rm tot}},n,b/a)$ in a similar way as described in \myrefsec{sec:corrfunction}. Going back to \autoref{fig:calib}, the measurements with the calibration function applied are shown in the filled and hatched histograms on the right two panels (for different magnitudes and $n$). Furthermore, the left panels show the 1-to-1 comparison of the size measurements with a running median with $1\sigma$ scatter (dashed). The comparison of the fully calibrated sizes with the HST based size measurements show that we are able to recover $R_{e}$ on UltraVISTA to an accuracy of better than $50\%$ ($1\sigma$ scatter).
	As shown in Figure~\ref{fig:calib}, the uncertainty of the calibrated sizes of galaxies close to the resolution limit of UltraVISTA can be up to a factor of three. We note that less than $5\%$ of our massive $\logm > 11.4$ are unresolved and thus could have much larger uncertainties.

\begin{figure*}[t!]
\begin{center}
\includegraphics[width=0.9\textwidth, angle=270]{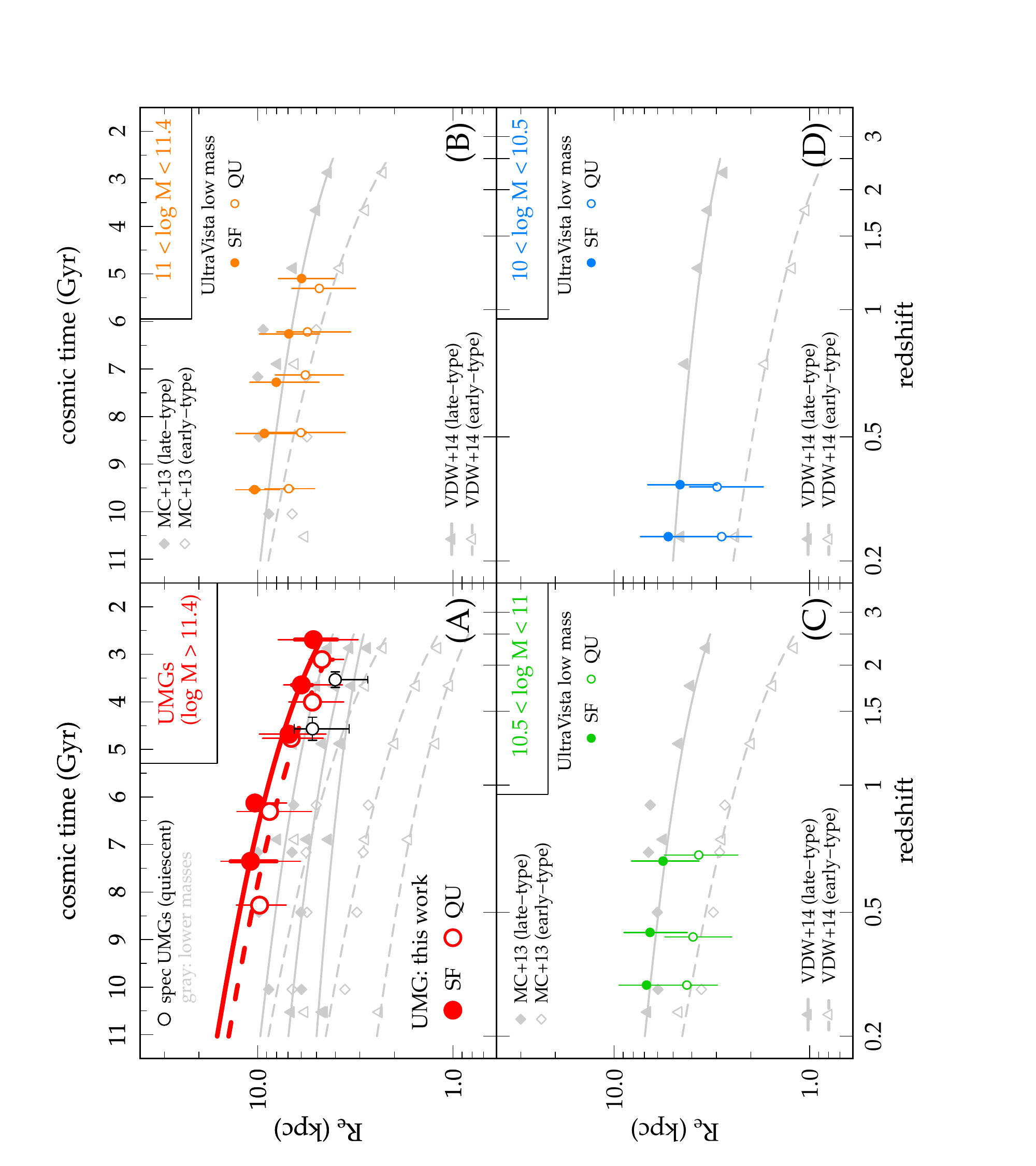}
\caption{
Size evolution as a function of redshift of star-forming and quiescent galaxies at different masses.
\textbf{Panel (A)} -- For quiescent (open, red) and star-forming (filled, red) UMGs at $\logm > 11.4$ with fits (dashed, solid red lines). Spectroscopically confirmed quiescent UMGs from the literature \citep{KROGAGER14,ONODERA14,BELLI15} are shown as black circles (median in two redshift bins with scatter). Clearly, star-forming UMGs are systematically offset to larger observed sizes at all redshifts. However, this offset is smaller compared to lower masses (shown in gray from literature). The thin error bars show the $1\sigma$ scatter and the thick error bars show the error on the median.
\textbf{Panels (B) through (D)} -- Comparison of our size measurements (color points, up to redshift where mass complete) with measurements in the literature in gray for $11.0 < \logm < 11.4$, $10.5 < \logm < 11.0$, and $10.0 < \logm < 10.5$, respectively. The lines (dashed: quiescent, solid: star-forming) are fits to the size evolution by \citet{VDW14}. The filled (open) symbols show measurements of star-forming (quiescent) galaxies from \citet[][]{CAROLLO13} (diamonds) and \citet[][]{VDW14} (triangles). The good agreement to our measurements shows that our measurement are not biased.
\label{fig:size}}
\end{center}
\end{figure*}

\subsection{Correction for internal color gradients}

Mostly negative internal color gradients are ubiquitously measured in star-forming galaxies up to at least $z\sim3$, whereas this effect is much less strong in quiescent galaxies \citep{CASSATA10,BOND11,SZOMORU11,WUYTS12,CIBINEL13a,PASTRAV13,BOND14,HEMMATI14,VULCANI14}. The observed color gradients are caused by different stellar populations and dust attributed to inside-out growth of galaxies and therefore depend on galaxy age, stellar mass, redshift, and star-formation activity. Such color gradients cause the observed size to change as a function of wavelengths. Vice versa, at a fixed observed wavelength the observed size of galaxies changes as a function of redshift since the rest-frame wavelengths shifts. The effect of color gradients may introduce artificial effects in the size evolution across redshift.
	Several studies have constrained this effect using observations at different wavelengths for different types of galaxies and stellar masses at various redshifts \citep[e.g.,][]{KELVIN12,VDW14,LANGE15}. Typical gradients for galaxies at $\logm = 10$ are on the order of $\left| \Delta \log R / \Delta \log \lambda \right| = 0.1 - 0.3$ depending on data quality, resolution, and redshift. This leads to corrections in size of $10-50\%$ over a wavelength range of rest-frame $0.5-1.0\,\mu {\rm m}$. In the following, we use the parameterization by \citet[][]{LANGE15} to correct our size measurements for internal color gradients. However, other parametrization \citep[e.g.,][]{VDW14} result in similar corrections and do not change the results of this paper.

\subsection{Verification of accuracy of size measurement}

Because our measurements at $\logm > 11.4$ are unique so far, we cannot directly check whether these are reasonable.

In the following, we use our (fully calibrated and mass complete) low-mass control samples at $10.0 < \logm < 11.4$ (see \myrefsec{sec:galselection}) to investigate possible systematics in our size measurement.

	In panels B through D of \autoref{fig:size} we compare our measured size evolution of quiescent (open, color) and star-forming (filled, color) galaxies to measurements taken from the literature \citep[gray lines and symbols;][]{CAROLLO13,VDW14}. The latter are based on high-resolution HST imaging and corrected for color gradients in the same way as we do here. We find a very good agreement with our measurements.
	
	In panel A we compare our final size evolution at $\logm > 11.4$ to spectroscopically confirmed quiescent galaxies at the same stellar mass in two redshift bins from the literature as black circles \citep[][]{KROGAGER14,ONODERA14,BELLI15}. These galaxies reside well within the $1-2\sigma$ scatter of our measurements (indicated by the thin error bar), although at the lower end. This can be explained by the higher success rate of spectroscopic surveys for compact galaxies with high surface brightness.

	Concluding, we do not expect any severe systematic biases in our measurements.

\begin{figure*}[t!]
\begin{center}
\includegraphics[width=0.42\textwidth, angle=270]{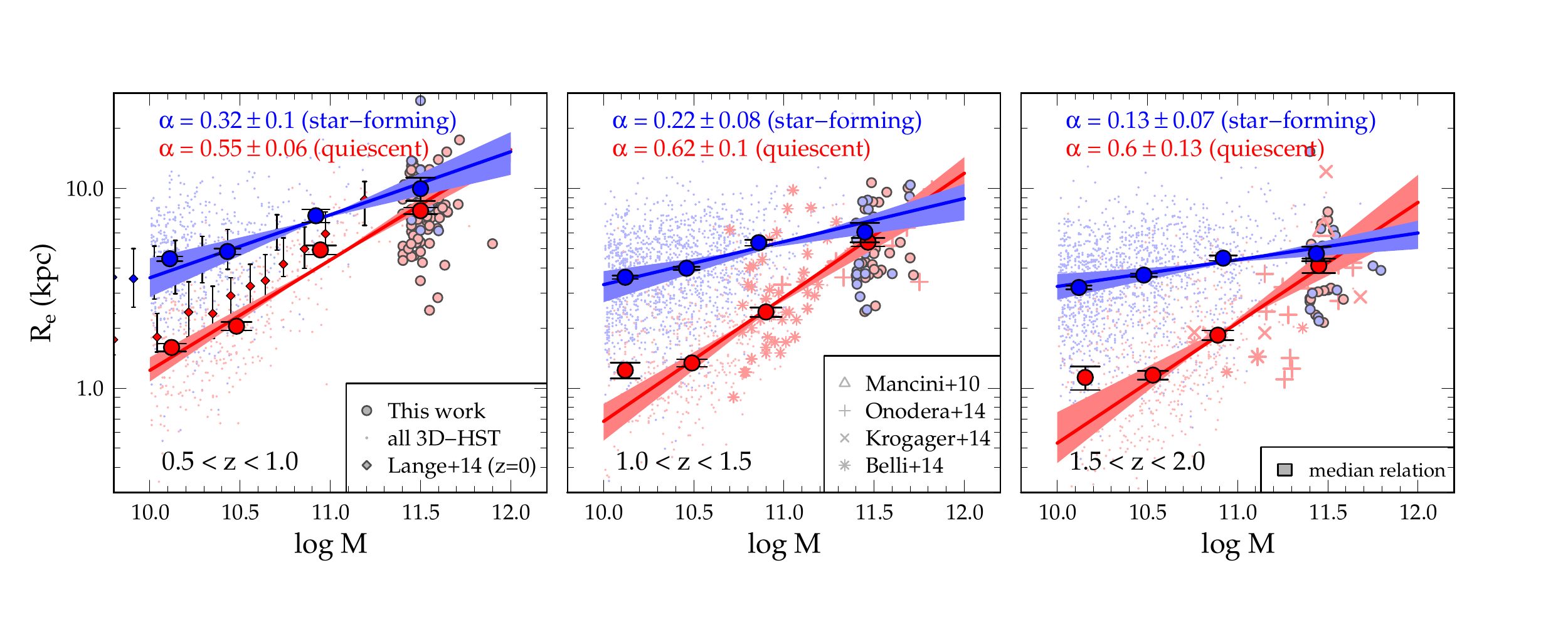}
\caption{
Mass versus size relation for quiescent (red) and star-forming (blue) galaxies for the combined sample of 3D-HST at $\logm \lesssim 11.5$ (thin dots) and our sample of massive galaxies with $\logm > 11.4$ (large points). Also included are spectroscopically confirmed quiescent galaxies from \citet[][pluses]{ONODERA14}, \citet[][crosses]{KROGAGER14}, and \citet[][asterisks]{BELLI15} at various masses as well as $z<0.1$ galaxies from \citet[][small dots with error bars]{LANGE15}. The medians in mass bins (large points with error bars) are fit linearly in log-space for three redshift bins using a least-square method (red and blue points with error bars indicating error on the median).  The logarithmic slope $\alpha$ ($R_{e} \propto m^{\alpha}$) is indicated for star-forming and quiescent galaxies. We note the large dispersion of sizes at high masses, consistent with the findings of \citet{MANCINI10}.
\label{fig:masssize}}
\end{center}
\end{figure*}

\section{Results: Size evolution of very massive galaxies} \label{sec:results}

\subsection{Size evolution of massive galaxies}

In \autoref{fig:size} (panel A) we show the final median size evolution with cosmic time of our massive $\logm > 11.4$ quiescent (red, open) and star-forming (red, filled) galaxies. These are compared to literature measurements at lower masses \citep[][gray lines and symbols]{CAROLLO13,VDW14} and spectroscopically confirmed quiescent galaxies at $z>1$ \citep[][median in two redshift bins, black open points]{KROGAGER14,ONODERA14,BELLI15}. 
	The dashed and solid line show fits to the size evolution of quiescent and star-forming galaxies, respectively, parametrized as $R_{e} = B\times (1+z)^{-\beta}$. We find a slope $\beta=1.22\pm0.20$ and $\beta=1.18\pm0.15$ for quiescent and star-forming galaxies with $\logm>11.4$, respectively. Note that this slope is statistically identical in contrast to lower masses where quiescent galaxies show a faster size increase with cosmic time than star-forming galaxies on average. Also, very massive star-forming galaxies are only $\sim20\%$ larger on average at a fixed redshift and stellar mass, whereas at lower masses the different can be as large as a factor of two (see gray lines).

\subsection{The stellar mass vs. size relation}
The relation between stellar mass and size (MR relation) has been measured so far on statistically large samples at $\logm < 11.0$. Our measurement on a large sample of galaxies at $\logm > 11.4$ enables us to provide an additional data point at high masses.
	In \autoref{fig:masssize}, we show the MR relation in three redshift bins measured over two orders of magnitudes in stellar mass. Shown are our data at $\logm>11.4$ (large filled dots) for quiescent (red) and star-forming (blue) galaxies as well as measurement at lower masses. The latter include the 3D-HST survey \citep[cloud of thin points,][]{VDW14}, spectroscopically confirmed quiescent galaxies at $z > 1$ \citep[crosses, asterisks, and pluses,][]{KROGAGER14,ONODERA14,BELLI15}, and galaxies at $z<0.1$ with measurements in $g$-band from the Galaxy and Mass Assembly survey \citep[small points with error bars,][]{DRIVER11,LANGE15}. The large blue and red symbols show the median size of star-forming and quiescent galaxies in different stellar mass bins. The lines show the corresponding log-linear fits ($R_{e}(m) \propto m^{\alpha}$, see \autoref{tab:fits2}) to the medians with errors from bootstrapping.

\begin{deluxetable}{cccc}
\tabletypesize{\scriptsize}
\tablecaption{Power-law slope ($R_{e}(m) \propto m^{\alpha}$) of the stellar mass vs. size relation at $z>0.5$ (this work including 3D-HST and spectroscopically confirmed quiescent galaxies) as well as integrated over cosmic time at lower redshifts (see references).
\label{tab:fits2}}
\tablewidth{0pt}
\tablehead{
\colhead{redshift range} & \colhead{star-forming} & \colhead{quiescent} & \colhead{Reference} \\
\colhead{ } & \colhead{$\alpha_{{\rm sf}}$} & \colhead{$\alpha_{{\rm qu}}$} & \colhead{}
}
\startdata\\
$ z \sim 0$ & $0.14~{\rm to}~0.39$ & $0.56$ & (1)\\
$ z < 0.1$ & $0.19 \pm 0.02$ & $0.41 \pm 0.06$ & (2)\\
$0.5 < z < 1.0$ & $0.30 \pm 0.10$ & $0.55 \pm 0.05$ & This work\\
$1.0 < z < 1.5$ & $0.22 \pm 0.08$ & $0.62 \pm 0.09$ & This work\\
$1.5 < z < 2.0$ & $0.14 \pm 0.06$ & $0.59 \pm 0.15$ & This work\\
$ z < 3 $ & $0.22 \pm 0.05$ & $0.75 \pm 0.05$ & (3)\\
\enddata
\tablenotetext{~}{(1) \citet{SHEN03}. For star-forming galaxies they fit $\alpha=0.14$ at $\logm < 10.6$ and $\alpha=0.39$ at $\logm > 10.6$.}
\tablenotetext{~}{(2) \citet{LANGE15}}
\tablenotetext{~}{(3) \citet{VDW14}. Report no significant change in slope over $0 < z < 3$.}
\tablenotetext{~}{~}
\end{deluxetable}


\begin{figure*}
\centering
\includegraphics[width=2.1\columnwidth, angle=0]{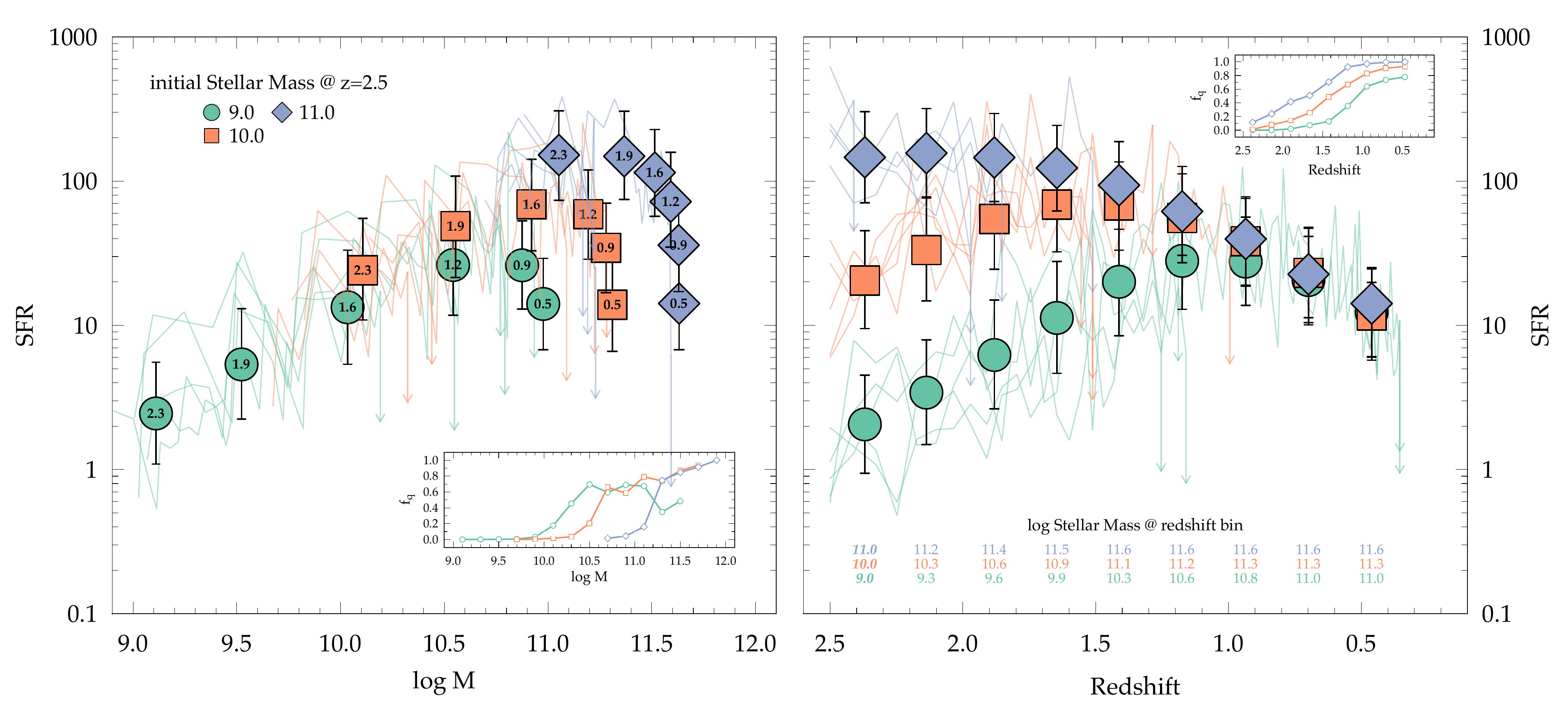}
\caption{
Tracks of our model galaxies on the SFR vs. stellar mass (left) and SFR vs. redshift (right) plane. We show the evolution of three sets of galaxies with different initial stellar masses at $z=2.5$ ($\logm = 9$, 10, and 11 shown in green circles, orange squares, and blue diamonds, respectively). The big symbols show the SFR in redshift bins with $\Delta z \sim 0.3$. The redshifts are indicated as numbers in the symbols on the left panel and the median stellar mass in each redshift bin is shown as numbers in the lower part of the right panel. The error bars represent $1\sigma$ errors.  The lines show five randomly chosen individual galaxies in each of the sets with different initial masses.
The insets show the quiescent fraction $f_q$ as a function of stellar mass marginalized over all redshifts (left panel) and as a function of redshift marginalized over all stellar masses (right panel).
\label{fig:visualization}}
\end{figure*}

The MR relation of quiescent galaxies is much steeper than for star-forming galaxies. The average sizes of quiescent and star-forming galaxies are comparable at $\logm \sim 11.5$ independent of redshift. The logarithmic slope of the MR relation ($\left<\alpha_{{\rm qu}}\right> \sim 0.6$ for quiescent and $\left<\alpha_{{\rm sf}}\right> \sim 0.2$ for star-forming galaxies) does not evolve significantly with cosmic time. Also, it is very consistent with the measurements in the local universe \citep[$z<0.1$;][]{SHEN03,LANGE15} finding values between $\alpha_{{\rm qu}} \sim 0.35-0.60$ and $\alpha_{{\rm sf}} \sim 0.15-0.25$ for quiescent and star-forming galaxies, respectively (see also \autoref{fig:masssize}). The study by \citet{VDW14} finds steeper slopes for quiescent galaxies ($\alpha \sim 0.75$), likely due to missing very massive quiescent galaxies at $\logm>11.5$.
\textit{The constant slope of the MR relation is indicative of a constant relation between the growth of galaxies in size (e.g., due to accretion) and stellar mass over cosmic time.}
		
	Finally, we note that a recent study by \citet{PENG15} suggests that the bulk of star-forming $z\sim0$ galaxies at $\logm < 11$ are being quenched via strangulation\footnote{Strangulation means the supply of cold gas is halted and thus star formation is shut down.} within $\sim4~{\rm Gyrs}$. We would therefore expect the $m-R_e$ relation of star-forming galaxies at $z\sim0.5$ and $\logm<11.0$ to be similar to the relation of the quiescent galaxies at $z\sim0$ if the observed sizes of the galaxies do not change during or after quenching. This is, however, not seen from the left panel of \autoref{fig:masssize} showing that star-forming galaxies $\sim4~{\rm Gyrs}$ ago are significantly (up to factor two) larger than local quiescent galaxies at $\logm<11.0$. This ``tension'' can be alleviated by post-quenching disk-fading, which would substantially decrease the observed sizes of quiescent galaxies and is shown to be at work at low redshifts \citep[][]{CAROLLO14} and most likely also at $z\sim2$ \citep[][]{TACCHELLA15b,TACCHELLA16b}.  In addition to this, at high redshifts, morphological transformation as a result of quenching cannot be ruled out.

\section{Model for the size evolution of massive quiescent galaxies}
\label{sec:model}

	The similar sizes at a given redshift of star-forming and quiescent galaxies at $\logm>11.0$ at all redshifts $z<2$ suggest a very close connection of these galaxies. This might have important implications on the process that quenches these galaxies.
	In this section, we investigate this further by modeling the size evolution of quiescent galaxies thereby applying different assumptions on the quenching process.

\subsection{Evolution of star-forming model galaxies}	

	Our model assumes that galaxies -- as long as they are forming stars -- evolve along the star-forming main-sequence (MS) spanned by stellar mass and SFR. In addition, we assign our model galaxies a half-light radius $R_e$ and a gas fraction $f_{\rm gas}$ using empirical relations and observations. The galaxies get eventually quenched in concordance with the observed quiescent fraction observed as a function of redshift and stellar mass.
	As fiducial model for quenching we assume an instantaneous (on-the-spot) quenching process without any change in the structure (i.e., half-light size $R_e$) of the galaxies. We complement this model with two additional models featuring a structural change (compaction due to starburst) as well as a delayed quenching. These different assumptions of quenching processes are explained in more detail later on.
	
	The main steps of our empirical model are the following.
	
\begin{figure}[t!]
\begin{center}
\includegraphics[width=1.1\columnwidth, angle=0]{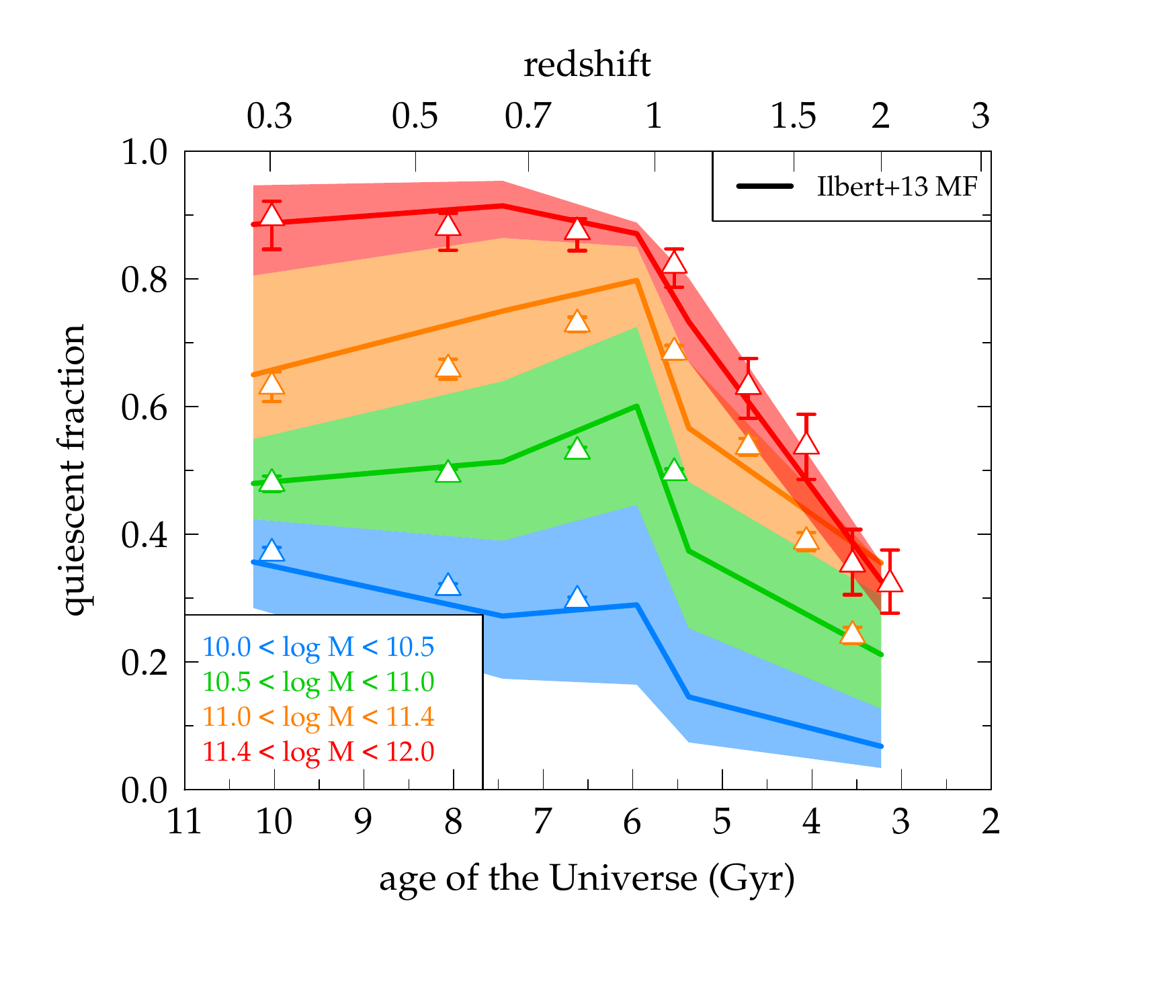}
\caption{Fraction of quiescent galaxies, $f_{q}(m,z)$, as a function of redshift for different bins in stellar mass (color). The lines show continuous fractions derived from the ratio of the stellar mass function of quiescent and star-forming galaxies from \citet{ILBERT13}. The symbols show the fraction derived from mass complete samples in UltraVISTA.
\label{fig:fq}}
\end{center}
\end{figure}

	\begin{itemize}
	\item Our model starts at $z = 2.5$ and uses the observed stellar mass function by \citet{ILBERT13} as initial condition for the mass distribution of the $100,000$ simulated star-forming galaxies with stellar masses between $7 < \logm < 12$. The initial mass distribution and fraction of quiescent galaxies is derived from the quiescent fraction $f_{q}(m,z)$ at a given redshift and stellar mass (\autoref{fig:fq}).
	
	\item We evolve the stellar mass and SFR of star-forming galaxies along their MS, for which we use the parameterization by \citet{SCHREIBER15} compiled from deep Herschel observations. We verified that the use of other parameterizations of the MS does not change the results and conclusions of this work. Furthermore, we assign to each of the star-forming model galaxies gas fractions $f_{\rm gas}(m,z)$ from our compilation of literature as outlined in \autoref{app:gasfraction} as well as sizes according to the measured size vs. stellar mass relation $R_e(m,z)$ for star-forming galaxies (including our new measurements at $\logm > 11.0$). When drawing values from the above empirical relations, we also include the observed scatter, which we characterized by a gaussian centered on the median. The typical scatter for the MS and MR relation is assumed to be $\sim0.3\,{\rm dex}$.
	
	\item At each redshift, we quench galaxies in mass bins randomly such that the model quiescent fraction reproduces the observed $f_{q}(m,z)$. After a galaxy is quenched, we set its gas fraction and SFR to zero. The remaining gas is added instantaneously to the stellar mass under the simple assumption that the gas is fully converted into stars and not stripped. Depending on the quenching model (see below), the new stars are either distributed evenly in the galaxy's disk or in a central region of $1\,{\rm kpc}$. We also do not implement the rejuvenation of galaxies once they are quiescent.
		
	\end{itemize}

For visualization purposes, example tracks of our model galaxies with different initial stellar masses at $z=2.5$ as well as the fraction of quiescent galaxies ($f_q$) are shown in \autoref{fig:visualization}.

\subsection{Quenching of model galaxies}

	We implement three simple models that should bracket different pathways of quenching processes. In the following, we describe these in more detail.

\textbf{Instantaneous / no structural change.} This is our fiducial model in which galaxies quench instantaneously without any structural change. A physical scenario could be the cut off of cold gas inflow by the heating of the gas in massive dark matter halos above $10^{12}\,\Msol$ \citep[e.g.,][]{CROTON06}. The galaxy then consumes its remaining gas according to its SFR on the star-forming MS and increases its mass evenly in its disk. 

\textbf{Instantaneous / compaction.} In this model, the galaxy decreases its overall size (compaction) due to an increase of its surface density instantaneously after the shut down of star formation. We assume that this compaction is triggered by a starburst in the inner $1\,{\rm kpc}$ region of the galaxy, which may be induced by a major merger event \citep[e.g.,][]{BARRO13}.
We compute the decrease in overall half-light radius after the starburst by adding the gas of the galaxy in a $1\,{\rm kpc}$ bulge component characterized by a $n=4$ Sersic profile to the disk dominated ($n=1$) star-forming galaxy.
For simplicity we assume that all of the gas mass is turned into stars in the bulge component. Furthermore we assume that the bulge component has the same mass-to-light ratio as the disk such that the ratio in luminosity of the bulge component and the disk is proportional to the ratio of stellar mass added to the bulge and stellar mass in the disk.

\textbf{Delayed / no structural change.} This model is similar to our fiducial model, however, the quenching does not happen instantaneously, but with a delay. A possible scenario could be the slow consumption of gas off the star-forming MS after the gas supply onto the galaxy is cut off. We assume the delay (i.e., the time the galaxy spends in the green valley) to be $50\%$ of the cosmic time between the start of quenching and $z=2.5$ ($\sim800\,{\rm Myrs}$ at $z=1.5$ and $\sim3\,{\rm Gyrs}$ at $z=0.5$).

\begin{figure*}[t!]
\centering
\includegraphics[width=2\columnwidth, angle=0]{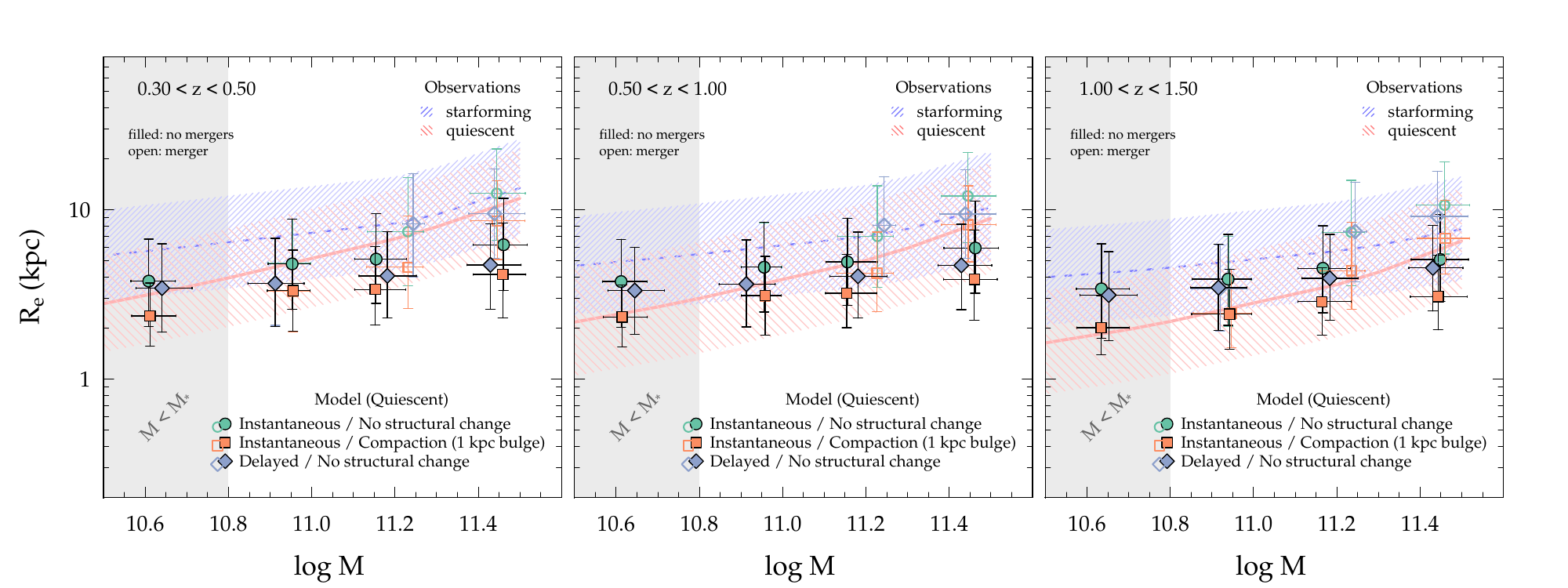}
\caption{
The observed MR relation in three redshift bins of star-forming (blue, hatched) and quiescent (red, hatched) galaxies together with our three different quenching models (symbols). The filled symbols show the models without mergers and the open symbols assume that $90\%$ of the quiescent galaxies above $\log(M_*) \sim 10.8$ experience ten $1:10$ minor mergers after being quenched. The models suggest a fast quenching of massive ($\logm > 11$) galaxies at $z>1$. In order to reproduce the MR relation at lower redshifts, a series of minor mergers is necessary. The same is true for quenching and compaction via major mergers. 
\label{fig:results}}
\end{figure*}

\section{Discussion}\label{sec:discussion}

	We now compare the predicted size evolution of quiescent galaxies from our simple empirical models with observations to investigate possible processes that quench massive galaxies at $z<2$.
	\autoref{fig:results} shows the MR relation of our quiescent model galaxies (symbols) together with the observed relations for star-forming (blue - hatched and dashed line) and quiescent (red - hatched and solid line), respectively. The width of the hatched bands and the error bars on the points represent the scatter in the observe as well as modeled relations. The different quenching models are shown in different colors and symbols as indicated in the legend.
	
	Focusing on galaxies above the characteristic knee of the stellar mass function \citep[$\log(M_*)\sim10.8$, e.g.,][]{ILBERT13,DAVIDZON17} we note the following.
	
	\begin{itemize}
	
	\item[\textit{(i)}] The instantaneous quenching model without altering the structure of the galaxies (green circles) predicts the quiescent MR relation well at all masses at $z>0.5$. However, this model under-predicts the sizes of the most massive galaxies ($\logm \sim 11.5$) at $z<0.5$.

	\item[\textit{(ii)}] The instantaneous quenching model followed by a compaction triggered by a starburst within a $1\,{\rm kpc}$ central region (orange squares), predicts well the sizes of $\logm<11$ galaxies down to $z\sim0.5$, but under-predicts the sizes at later times as well as at higher masses.

	\item[\textit{(iii)}] The delayed quenching model without structural change (blue diamonds) is only able to explain the MR relation at $\logm > 11$ at the highest redshifts, but under-predicts the sizes at lower redshifts by factors of $2-3$. It reproduces well the relations at $\logm<11$ and $z<1$.
	
	\end{itemize}
	
	We explain and interpret these findings in more detail in the following sub-sections.

	\subsection{Slow versus fast quenching at $m>M^{*}$}
	
	The star formation in very massive galaxies can be shut down without significant structural change of the light profile by cutting off the gas supply onto the galaxies. In current theoretical models and simulations, this can be achieved in galaxies with massive dark matter halos of $m_{\rm DM} > 10^{12}\,\Msol$, which cause the infalling gas to be heathen up \citep[e.g.,][]{CROTON06}. This may result in a uniform decrease of the star formation in the galaxy's disk without altering its structure significantly.
	Taking the above results at face value suggests that if there is no net structural change after the turn-off of star formation, massive galaxies ($\logm > 11$) have to transition from star-forming to quiescent on relatively short time scales. This is suggested by the fact that our fiducial model (instantaneous quenching) is able to reproduce the sizes of galaxies at these stellar masses reasonably well, at least in the two upper redshift bins at $z\gtrsim0.5$.
	An instantaneous quenching might be too much of a simplification and a non-zero quenching time is suggested by recent observational studies \citep[e.g.,][]{SCHAWINSKI14,PENG15}. Our delayed quenching model works well for redshifts $z\gtrsim1$ and $\logm>11$, where the delay times are  shorter than $1-1.5\,{\rm Gyrs}$ according to our definition ($50\%$ of the difference in cosmic time between the quenching event and $z=2.5$). Note that this is compatible with the time a galaxy on the star-forming MS needs to consume all of its gas given its main-sequence gas fraction and SFR: less than $1-2\,{\rm Gyrs}$ for a galaxy at $\logm>11$ and $z>0.5$ \citep[e.g.,][]{TACCONI17}.
	Note that our delayed model over-predicts the sizes of galaxies at $m< M^{*}$ at $z>1$. This would suggest that the delay as defined here is not long enough and instead a longer delay ($2\,{\rm Gyrs}$ or more) is favored. This mass dependence of the quenching time (i.e., slow vs. fast quenching) is also strongly suggested by recent simulations \citep[e.g.,][]{HAHN16} and could be explained by different quenching processes taking place at different stellar masses and as a function of environment the galaxies are living in.

	\subsection{Merger-induced starbursts and compaction}
	
	It is suggested that mergers play an important role in shaping galaxies at high stellar masses. Thus a smooth quenching without significantly altering the structure of a galaxy is likely too simplistic. Our model of a merger-triggered compact starburst inducing a fast consumption of gas and quiescence therefore might be a better approach to characterize the quenching mechanism at high redshifts and high stellar masses.
	As shown in Figure~\ref{fig:results}, such a scenario under-predicts the sizes of massive quiescent galaxies by factors of two or more at all redshifts. If such a scenario is the dominant way of quenching massive galaxies, then the galaxies have to grow individually to meet the observed MR relation. This is similar to the fast-track quenching mechanism proposed by \citet{BARRO13} \citep[see also][]{ZOLOTOV15} in which galaxies experience a compaction phase with subsequent growth due to minor and major mergers.
	Note that changing the parameters of this particular model does not significantly change this conclusion. For example, assuming a $2\,{\rm kpc}$ central starburst would only increase the sizes by $\sim50\%$ and still leads to a significant under-prediction.

	\subsection{Post-quenching growth through mergers in massive galaxies}
	
	Figure~\ref{fig:results} shows that \textit{all} of our bracketing models in some cases severely under-predict the sizes of quiescent galaxies above $M^*$ and $z\lesssim1$. One possible way to bring the models in agreement with observations is to introduce a series of minor and/or major mergers following the quenching event. 
	We investigate this further by assuming a simple toy model in which $90\%$ of the quiescent galaxies above $\log(M_*) \sim 10.8$ experience ten $1:10$ minor mergers during their lives after being quenched. We choose this case because minor mergers are more common and are dominantly increasing the size of galaxies and less their stellar mass.
	For the implementation of this model, we assume that the virial condition holds for gas-poor ellipticals and compute the resulting size increase ($\Delta R_e$) as a function of the merger mass fraction ($\Delta m$) and change in velocity dispersion ($\Delta \sigma$) during the merger event as
	
	\begin{equation}\label{eq:vir}
	\Delta R_e = \Delta m \left(\frac{1}{\Delta \sigma} \right)^{2},
\end{equation}
	
	where $\Delta R_e = \frac{R_{e,post}}{R_{e,pre}}$, $\Delta m = \frac{m_{post}}{m_{pre}}$, and $\Delta \sigma = \frac{\sigma_{post}}{\sigma_{pre}}$ are the ratios of quantities before (``pre'') and after (``post'') the merging event. We assume that the change in the velocity dispersion is negligible during the merger event, i.e., $\Delta\sigma \sim 1$ \citep[e.g.,][]{HOPKINS09a,OSER12}.
	
	The open symbols in \autoref{fig:results} show the impact of post-quenching mergers on our previous results. The addition of a series of minor mergers to our instantaneous quenching $+$ compaction model (orange open squares) leads indeed to a good agreement with the observed MR relation at $z>0.5$ at all stellar masses probed here. We note, however, that the sizes of massive $m>M^*$ galaxies are still under-estimated at $z<0.5$. It is therefore likely that, if the compaction model holds, these galaxies must experience more minor mergers cumulatively than anticipated in our simple merger toy model. Alternatively, massive galaxies at later cosmic times might be quenched via other paths that do not include a compaction phase, such as heating of cold gas alone. Such a possibility is shown by our fiducial model with post-quenching minor mergers (green open circles), which is able to predict the sizes of massive galaxies at $z<0.5$ well (however, fails at high redshifts).
	We note that Equation~\ref{eq:vir} describes the effect of size growth by major mergers. Instead, strictly speaking, the size growth due to minor mergers is expected to be steeper \citep[$\Delta R \propto \Delta m^{\alpha}$ with $\alpha>1$, e.g.,][]{BEZANSON09,NAAB09}, which would decrease the number of mergers needed in our model. For example, assuming $\alpha=2$, we find that only $\sim30\%$ of the galaxies at $\logm>10.8$ are needed to experience a $1:10$ merger in order to meet the observations.

	Finally, we note that mergers for low-redshift ($z<1$) less-massive ($m<M^{*}$) are not needed to bring our models in agreement with observations. This is in line with the idea that the size evolution of quiescent galaxies below $\logm\sim11$ with cosmic time is mainly driven by the addition of newly quenched galaxies, while at higher masses it is more dominated by individual growth due to mergers \citep[e.g.,][]{CAROLLO14,BELLI15}.
	

\section{Summary \& conclusions}\label{sec:ending}

We use the size evolution of massive star-forming and quiescent galaxies as an independent diagnostic tool to investigate the process of quenching at $\logm>11$ and $z\lesssim2$.
	To this end, we measure the half-light size evolution of a large sample of very massive star-forming and quiescent galaxies at $\logm~\gtrsim~11.4$ on the 2-square degree survey field of COSMOS/UltraVISTA. We find the size evolution of both populations of galaxies at $\logm~>~11.4$ to be similar in slope and normalization and to be consistent with the extrapolation of the mass versus size relation from lower masses.
	
	In order to investigate different quenching mechanisms and the impact of mergers, we predict the MR relation of massive $m>M^*$ quiescent galaxies within our simple empirical models as a function of redshift.
	Our main results are the following.
	
	\begin{itemize}
	
	\item Massive galaxies quench fast. Models with instantaneous quenching or short delay of up to $\sim1\,{\rm Gyr}$ are able to predict the sizes of quenching galaxies at $z>1$ and $m>M^*$. Longer quenching times are more favored at lower masses and redshifts.
	
	\item A more realistic model incorporating a compaction phase (e.g., due to merger-triggered central starburst within $1\,{\rm kpc}$) followed by quiescence and subsequent individual growth by mergers is able to reproduce the observed MR relation of massive $m>M^*$ quiescent galaxies at all redshifts. 
		
	\item None of our models is able to predict the size evolution of $m>M^*$ galaxies at low redshifts ($z\lesssim1$). We show that with $1:10$ minor mergers for $90\%$ of the quiescent galaxies at $m>M_*$ the models can be brought into agreement with observations. In contrast, no mergers are needed at lower stellar masses in agreement with the size evolution being driven by the addition of bigger, newly quenched galaxies.
	
	\end{itemize}
	
	It is important to note that we are not able to distinguish the dominant pathways of quenching of massive quiescent galaxies with our simple models as these yield very similar predictions for the size evolution. Nonetheless, our study suggests that quenching is likely a fast process at the stellar masses probed here with a significant involvement of mergers in the post-quenching growth of massive galaxies.
	For further distinguishing these models, more information on the (resolved) structural properties of the galaxies is necessary. This will be possible with high-resolution imaging and spectroscopy of massive quiescent galaxies by the HST or the \textit{James Webb Space Telescope}.

\acknowledgments

We would like to thank Dan Masters, Charles Steinhardt, Behnam Darvish, and Bahram Mobasher for valuable discussions. Furthermore, we would like to thank the referee for valuable feedback that greatly improved this manuscript.
AF acknowledges support from the Swiss National Science Foundation.
Based on data products from observations made with ESO Telescopes at the La Silla Paranal Observatory under ESO programme ID 179.A-2005 and on data products produced by TERAPIX and the Cambridge Astronomy Survey Unit on behalf of the UltraVISTA consortium.
This work is based on observations taken by the CANDELS Multi-Cycle Treasury Program with the NASA/ESA HST, which is operated by the Association of Universities for Research in Astronomy, Inc., under NASA contract NAS5-26555.


\bibliographystyle{aasjournal}
\bibliography{bibli2.bib}



\clearpage
\appendix

\section{The gas fraction $f_{{\rm gas}}(m,z)$} \label{sec:gasfraction}\label{app:gasfraction}

\begin{figure*}[t!]
\begin{center}
\includegraphics[width=0.8\textwidth, angle=270]{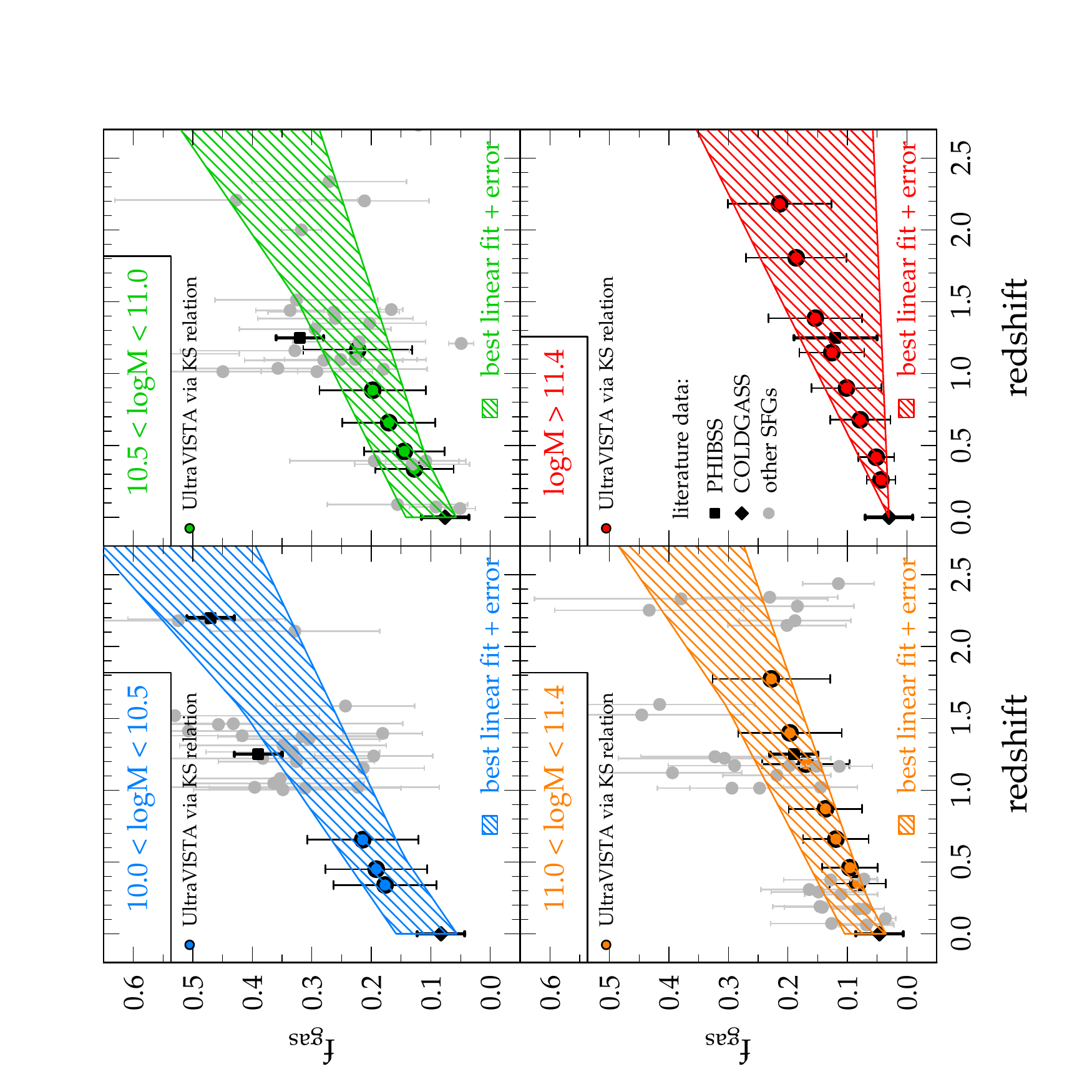}
\caption{Gas fraction as a function of redshift in the four different stellar mass bins (different panels).
The black and gray symbols show literature values measured from individual galaxies (black squares: PHIBSS at $z\sim 1-1.5$, \citet{TACCONI13}; black diamonds: COLDGASS at $z\sim0$, \citet{SAINTONGE11}; gray points: lensed and other star-forming galaxies from \citet{DESSAUGES14} and references therein). The hatched region is a fit ($f_{gas}(m,z)$) to the measured individual galaxies including uncertainty as described in the text.
For comparison, the color symbols show $f_{gas}$ derived from the Kennicutt-Schmidt relation for the UltraVISTA galaxies. Note that these are not used in the fitting.
\label{fig:gasfraction}}
\end{center}
\end{figure*}

We use studies from the literature to fit an empirical relation $f_{{\rm gas}}(m,z)$, which is used in our models.
The data used include PHIBSS at $z\sim1-1.5$ \citep{TACCONI13} and COLDGASS at $z\sim0$ \citep{SAINTONGE11} as well as data from lensed and other star-forming galaxies from \citet{DESSAUGES14} and references therein.

The result is shown in \autoref{fig:gasfraction} for four different bins in stellar mass. The PHIBSS and COLDGASS data is shown in black, the other measurements are shown in gray.
We also show in color $f_{{\rm gas}}$ derived from our galaxies (UMGs and lower mass control sample) using the Kennicutt-Schmidt relation \citep[KS relation,][]{SCHMIDT59,KENNICUTT98}, relating $\Sigma_{gas} \propto \Sigma_{SFR}^{N}$, where we take $N=1.31$ \citep{KRUMHOLZ12}. Note that these derivations are not used for the fitting of the parametrization for $f_{{\rm gas}}(m,z)$.

We derive $f_{gas}(m,z)$ and its uncertainty ($95$\% CLs) by fitting the observed data as follows.
	We first perform a linear fit forced through the COLDGASS data point at $z=0$ in order to determine the slope. The error on the slope is derived from the systematic error of the fit and the uncertainty of the data points by bootstrapping which we both add in quadrature.
	In a second fit, we fix the slope to the one determined before and fit for the intercept including error.
The resulting uncertainty region as shown in \autoref{fig:gasfraction} as hatched region is then derived by the unification of the errors of the two fits.
To get a continuous function for $f_{gas}$, we interpolate between the four stellar mass bins.

	We compared our fit to the recent work by \citet{GENZEL15}. We find that their $f_{{\rm gas}}(m,z)$ parametrization has a slightly steeper redshift dependence resulting in $10-30$\% larger gas fractions at the highest redshifts. We have verified that our results do not change if using the \citet{GENZEL15} parametrization for $f_{{\rm gas}}(m,z)$.

\end{document}